\documentstyle[prb,aps,twocolumn,epsf]{revtex}

\def\prb{Phys.\ Rev.\ {\bf B}}

\def\prl{Phys.\ Rev.\ Lett.\/}

\def\jpc{J.\ Phys.\ {\bf C}}
\def\be{\begin{equation}}
\def\ee{\end{equation}}
\def\ba{\begin{eqnarray}}
\def\ea{\end{eqnarray}}

\def\C60{A$_x$C$_{60}$}

\begin{document}

\twocolumn[\hsize\textwidth\columnwidth\hsize\csname@twocolumnfalse\endcsname
\title
{Nematic phase of the two-dimensional electron gas in a magnetic field}
\author{Eduardo Fradkin$^{a}$, Steven A.~Kivelson$^{b}$, Efstratios 
Manousakis$^{c}$ and Kwangsik Nho$^{c}$}
\address{
$a)$ Department of Physics, University of Illinois, Urbana, IL 
61801-3080,
$b)$ Department of Physics, U.\ C.\ L.\ A.\  , Los Angeles, CA  
90095, and 
$c)$ Department of Physics and MARTECH, Florida State University, 
Tallahassee FL 32306-4350}
\date{\today}
\maketitle
\begin{abstract}
The two dimensional electron gas (2DEG) in moderate magnetic fields in
ultra-clean AlAs-GaAs heterojunctions exhibits transport
anomalies suggestive of a compressible, anisotropic metallic state.  
Using scaling arguments and Monte Carlo simulations, we
develop an order parameter theory of an electron nematic phase.
The observed temperature dependence of the resistivity anisotropy behaves like
 the orientational order parameter if the transition
to the nematic state occurs at a finite temperature, $T_c \sim 65 mK$, and
is slightly rounded by a small background microscopic anisotropy.  We
propose a light scattering experiment to measure the critical
susceptibility.
\end{abstract}

\

\

]

\narrowtext

Recently~\cite{paper1} two of us introduced the concept of liquid crystal
phases of the two-dimensional electron gas (2DEG) in large magnetic fields,
as an extention of earlier work on high temperature
superconductors~\cite{topo,nature}. Electronic liquid crystal phases are 
quantum mechanical analogues of classical liquid crystals,
and are predicted~\cite{topo,nature} to be a generic feature
of strongly correlated fermionic systems. In the case of the 2DEG the competing
effects of repulsive (Coulomb) interactions and the quenching of the kinetic energy
of electrons in Landau levels lead naturally to the existence of such
phases. Pursuing this analogy, the phases of a 2DEG 
in order of increasing symmetry breaking, were characterized as:
(a) isotropic liquids, (b) nematic liquids, (c) smectic
liquid crystals, and (d) insulating crystals.  The fluid character of
states (a) and (b) is obvious, as they are
translationally invariant. The smectic breaks transational symmetry 
in only one direction, and so is also a fluid.  Insulating 
crystals break the translational symmetry down to a discrete
subgroup, such that there are an integer number of electrons per unit cell.  

Among the isotropic liquids are
the various quantum Hall fluids, while insulating crystals
are simply the Wigner crystal and its generalizations. 
A smectic (or stripe) phase has been found in Hartree-Fock calculations, 
presumably accurate in high Landau levels~\cite{FPA,KFS,MC,SMP}, which
provides a qualitative picture of a smectic state. Moreover,
a recent exact diagonalization study of a system with 
12 electrons in the $N\ge 2$ Landau levels
found results consistent with a smectic (stripe) ground state~\cite{rezayi} 
(up to yet poorly understood finite size effects). Wigner crystals
and ``bubble phases" (crystalline states with several particles per unit
cell) have also been found as Hartree-Fock variational states.  In 
addition, a predicted charge-density-wave instability of the smectic phase at 
Hartree-Fock level\cite{paper1,nature,vadim,macdonald} leads to an 
insulating ``stripe-crystal'' phase with a rectangular unit cell.  
However, a microscopic theory of the nematic phase does not presently exist.

The low energy physics of a
quantum smectic can be understood in terms of a theory of quantum
fluctuations of the smectic~\cite{paper1,nature,vadim,macdonald}, 
including the phase transition to the stripe crystal phase.
The long-distance behavior of a quantum nematic phase is
completely determined by
symmetry and by the associated Goldstone modes.  Both 
smectic and nematic states break rotational symmetry, and as such 
transport properties of both states is anisotropic. In Ref.\ \ref{ref:paper1}
we argued that while the smectic and nematic states are both natural
candidates to explain the anisotropy observed in recent 
experiments~\cite{caltech,princeton} on 2DEG in  high mobility $Al As-Ga As$
heterostructures, at least at finite temperature, there are strong 
reasons to favor the nematic. The smectic 
has an infinite conductivity, at least in one direction,
whereas the measured 
(anisotropic) conductivity has a finite $T \to 0$ limit.
 In addition, the data shows a pronounced temperature
dependence of the resistivity,
consistent with the existence of a finite temperature phase transition; 
since the energy of a dislocation is still finite even for a Coulomb interaction
the 
smectic always melts at {\it any} non-zero temperature~\cite{2Dsmectics}. 
In the present paper we explore the universal properties of 
the 2D electron nematic.
 
The experiments of Refs.\ ~\ref{ref:caltech} and \ref{ref:princeton} have
revealed the existence of 
regimes of magnetic fields
in which the 2DEG exhibits characteristics of a compressible fluid with an 
unexpectedly 
large and temperature-dependent anisotropy in its
transport properties. This occurs when the
Landau level index $N$ lies in the range $2 \leq N \leq 6$. 
The large anisotropy is seen only in high mobility samples.
In the same samples,
similar behavior has also been seen in the first $N=1$ Landau level 
when the magnetic field is tilted~\cite{tilt}, while in 
the lowest $N=0$ Landau level,  
a large number of fractional quantum Hall (FQH) states are observed,
including a subtle FQH state at $\nu=5/2$. 
To summarize the experimental facts: (i)
Measurements on square 
samples show that, as the
temperature is lowered below $100$ mK, the longitudinal {\sl resistance}
$R_{xx}$ grows very rapidly while $R_{yy}$ becomes smaller; 
as $T\rightarrow 0$, their ratio approaches a constant~\cite{19b} in the range
$100 < R_{xx}/R_{yy} < 3,500$ 
where $x$ and $y$ are
orthogonal lattice directions. (ii)
In Hall bars~\cite{caltech,caltech-unpublished},
$R_{yy}$ is essentially
temperature independent while $R_{xx}$ increases by a factor of $5-10$ as the 
temperature is lowered below $100 mK$.
(iii) The compressible (dissipative) regime occupies a finite range of
magnetic fields $\Delta B$, centered around the middle of the partially
filled Landau level, and unlike the conventional transition between plateaus,
$\Delta B$ does not shrink as the temperature is lowered.
(iv) In the same range of magnetic fields the Hall resistance varies continuously 
 with the magnetic field. (v)
At the peak, the {\sl conductivity} $\sigma_{yy}$ 
is typically of the order of $e^2/h$ (see below). (vi) Applying an
in-plane magnetic field tunes the anisotropy through
zero, and reverses the roles of the $x$ and $y$ directions. (viii)
In the region of the resistivity peak, the $I-V$ curves are highly 
non-linear but do not show any threshold (depinning) behavior
~\cite{caltech,caltech-unpublished}. (ix)
There are reentrant integer quantum Hall plateaus symmetrically 
located for magnetic fields {\sl outside} the
compressible regime. (x)
The anisotropy has not been reported in lower mobility samples which show
instead the (expected) phase transition between quantum Hall states. 

A natural interpretation of the experiments is that, in regimes in which 
interactions 
dominate over the effects of disorder, instead of the expected transition between 
plateaus in the middle of the  partially filled Landau level, the 2DEG
forms a compressible anisotropic fluid. Because a continuous 
(rotational) symmetry
cannot be spontaneously broken in D=2,
for such an
anisotropy to be observable\cite{paper1} the sample must
have a small background microscopic anisotropy whose effect is greatly
amplified at low temperatures by the collective properties of the
state.
Specifically, we will show that the experimentally observed 
temperature dependence of the anisotropy 
can be understood as evidence for a finite temperature
Kosterlitz-Thouless (KT)
transition from a two-dimensional nematic to an isotropic 
fluid~\cite{2Dsmectics,KT}, which is
rounded by a symmetry breaking field representing the effects of the background
anisotropy. We present an analysis (Fig. 1) of the experimental data of 
Lilly {\it et al}~\cite{caltech}, 
and a fit with the results of a
Monte Carlo simulation of a model of a classical nematic in a symmetry
breaking field (Fig. 2).  The results
strongly support our
earlier claim~\cite{paper1} that the anisotropic transport occurs where
the 2DEG is in a nematic phase 
(at least at finite temperature).  They also give some
indirect support to the  further conjecture
made in Ref.\ \ref{ref:paper1}, that there is a direct transition
as a function of $B$ from the nematic state to an insulating
stripe crystal phase, which was identified with
the innermost of the
reentrant quantum Hall liquids.
(See, also Ref.\ \ref{ref:fertig}.)

In 2D, the nematic has only quasi-long-range order;
its finite temperature transition
to a disordered liquid can be described by the two-dimensional classical 
$XY$-model with a director order parameter~\cite{2Dsmectics}. 
Such a description should fail at (very) low
temperatures where quantum fluctuations (and/or quenched disorder) 
become important.
Since the order parameter of the nematic state is a director
field, $\vec m(\vec r)$, it is periodic under rotations by $\pi$, 
and has the form 
$m_{x}(\vec r)+im_{y}(\vec r)=\exp(2 i\theta_{\vec r})$.
The classical Hamiltonian of this system is thus
\begin{equation}
H=-J\sum_{{\vec r},\mu=x,y }\cos (2 \Delta_\mu \theta_{\vec r})+ h
\sum_{\vec r}  \cos (2 \theta_{\vec r})
\label{eq:XY} 
\end{equation}
where, for simplicity we have used a square lattice of unit spacing whose sites
are labelled by the lattice vectors $\vec r$. In Eq.\ \ref{eq:XY} we have use the
notation $\Delta_\mu \theta_{\vec r}=\theta_{{\vec r}+{\vec e}_\mu}-\theta_{\vec
r}$, where $e_\mu$ is a unit vector along the direction $\mu=x,y$, 
and $J$ is the stiffness, the energy required to rotate
two nearby regions by a small angle. The quantity $h$ breaks rotational symmetry
explicitly. It represents the effects of a background symmetry breaking field,
such as the anisotropy and/or the effects of a parallel magnetic field. 

Because two is the lower critical dimension for continuous symmetry 
breaking,
for $0 < T < T_c$ the system is controlled by a line of critical 
points~\cite{KT}. In this range of $T$, where
classical Goldstone
excitations of the nematic (``spin waves") dominate, the correlation
function $G(\vec r)=\langle \exp (2i[\theta(\vec r)-\theta(\vec 0)]\rangle$ 
of the order parameter has power law behavior, $G(\vec r) \sim 1/|\vec
r|^{\eta(T)}$, with  $\eta= 2T/\pi \kappa(T)$ and
a divergent susceptibility. Here $\eta \to 1/4$ as $T \to T_c$ and $\kappa(T)$
is the helicity modulus, which approaches $\kappa(0)=4J$
as $T \to 0$ and $\kappa(T_c)
=4T_c/\pi$. In the presence of a symmetry breaking field
$h$, the order parameter behaves like $m=\langle
\exp(2i\theta)\rangle\sim |h|^{1/\delta}$, where $\delta=4/\eta-1$ and
$\delta(T_c)=15$.
For $T>T_c$, the correlation length is finite, and diverges at $T_c$ like 
$\xi \sim \exp(A/{\sqrt{T-T_c}})$, where $A$ is a (non-universal) constant. 
At finite $h$ the correlation length is always finite, and the singularities 
of the KT transition get rounded.  In this regime, even a very small symmetry 
breaking field induces a very large expectation value of the order parameter.
For $T > T_{c}$, $ m \sim  \chi(T) h$, where $\chi(T)\sim \xi^{7/4}$ is the
susceptibility.

How is this thermodynamic transition, which describes the breaking of rotational
invariance, related to the transport properties? 
On general grounds one expects that near a
phase transition, quantities which transform in the same way under the symmetry
 should be related, even if one is a transport
coefficient and the other a thermodynamic property.
In particular~\cite{paper1} the combination of
resistivities $\zeta=(\rho_{xx}-\rho_{yy})/(\rho_{xx}+\rho_{yy})$ transforms like
the order parameter $m=\langle \exp(i 2\theta )\rangle$. It should be
related to the order parameter through an
odd {\sl analytic} function $\zeta=f(m)$.
Therefore, near $T_{c}$, if the symmetry breaking is
small, the linear approximation $f(m) \propto m+ O(m^3)$
should be reasonably accurate\cite{comment3}.

\begin{figure}
\begin{center}
\leavevmode
\vspace{.2cm}
\noindent
\epsfxsize=2.5in
\epsfysize=2.0in
\epsfbox{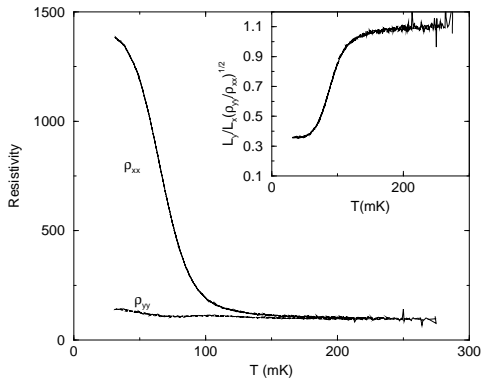}
\end{center}
\caption
{Resistivities $\rho_{xx}$ and $\rho_{yy}$determined from the resistance data
of Lilly {\it et al} [\protect\ref{ref:caltech}] 
[\protect\ref{ref:caltech-unpublished}] at
$\nu=9/2$, after deconvoluting the effects of the geometry; 
$\rho_{yy}$ is essentially constant for the entire range of
temperatures, as in Hall bars. Inset: the function $x(T)$.}
\label{fig:resistivities}
\end{figure}

We can determine if the 2DEG is in a nematic phase 
by analyzing the {\sl temperature}
dependence of the resistivity in terms of the temperature dependence of the
order parameter of the nematic in the presence of a symmetry breaking field.
What is needed is the function $m=\Phi(T,h)$, the equation
of state, which we  computed by a 
Monte Carlo simulation of the classical $XY$ model of Eq.\ \ref{eq:XY}. 
Notice that we relate $m$ to a local
(intensive) property such as the resistivity instead of to the resistance, which is
extensive and sensitive to significant finite size effects.
However, the experimental data gives the resistances $R_{xx}$ and
$R_{yy}$ as functions of temperature, not the resistivities. Thus, in order to 
fit the data, we extracted the resistivities from the measured
resistances, using a method discussed below, with 
the result shown in Fig.\ \ref{fig:resistivities}.

One result of this analysis is that
the  resistivities for the square sample behave exactly in
the same way as the resistances of the Hall bars.
Given
the low $T$ values of $\rho_{xx}$ and $\rho_{yy}$ in Fig. 1, and the
(large) measured value of the Hall conductance, one finds that the peak value of
the {\sl conductivity} is $\sigma_{yy} =1.12 e^2/h$ and $\sigma_{xx}=0.11 e^2/h$. 
Notice that $\rho_{xx}$
saturates rather sharply below $55 mK$ and that both $\rho_{xx}$ and $\rho_{yy}$
approach non-zero (and different) values as $T \to 0$. Thus, the 2DEG
remains in an  anisotropic compressible (metallic) state, down to the lowest
accessible temperatures. 
\begin{figure}
\begin{center}
\leavevmode
\vspace{.2cm}
\noindent
\epsfxsize=3in
\epsfysize=2.5in
\epsfbox{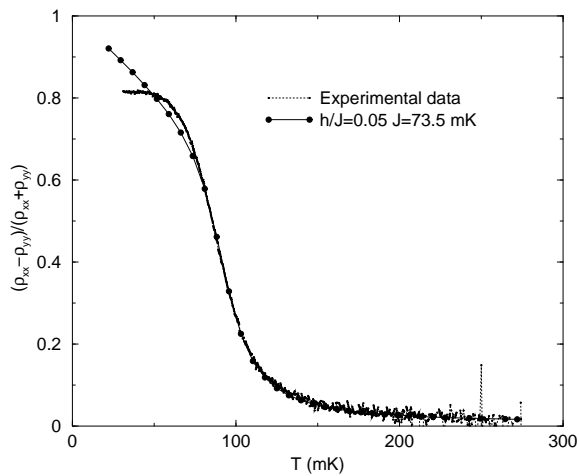}
\end{center}
\caption
{Fit of the Monte Carlo data for a $100 \times 100$ lattice,
to the data of Lilly {\sl
et.\ al.} [\protect\ref{ref:caltech}] [\protect\ref{ref:caltech-unpublished}]. The
best fit is found for $J=73 mK$ and $h=0.05 J=3.5 mK$ and $T_c=65 mK$. }
\label{fig:fit}
\end{figure}
Having determined the temperature dependence of the {\sl resistivities} we can now
see if it is consistent with a (rounded) phase transition from
a high temperature isotropic fluid phase to a low temperature nematic phase.
We have done this by means of a Monte Carlo Metropolis simulation of the 2D $XY$ model
of Eq.\ \ref{eq:XY} on square lattices of sizes $40 \times 40$ through $120 \times
120$, for the range of symmetry breaking fields $0.01J < h < 0.5 J$, and for a wide 
range of temperatures (see below). In Fig.\ \ref{fig:fit} we
show our Monte Carlo data for the order parameter as a function of temperature
for $h=0.05 J$. For this range of symmetry breaking fields, we find that 
for $L=100$ the finite size effects on the order parameter are very small. 
We have fitted
the data by assuming that $\zeta=(\rho_{xx}-\rho_{yy})/(\rho_{xx}+\rho_{yy})$
is actually {\sl equal} to the order parameter $m$.~\cite{caveat} 
Having done so, we
fitted the data by finding the best value of $J$ that fits the data for
a given value of $h$, and then changed $h$ to get the best fit. 

The classical nematic does indeed explain the temperature
dependence; the data is consistent with a thermodynamic Kosterlitz-Thouless
transition at $T_c(h=0)=0.88 J\sim 65 mK$ ~\cite{schultka},
slightly rounded by a background anisotropy field
of magnitude $h \sim 0.05 J \sim 3.5 mK$, which is a very small energy
scale. Notice that both the stiffness $J\sim 73 mK$ and $h$ are much smaller than the 
Coulomb energy, although they are comparable with the gap in the
$\nu=5/2$ (presumably paired) state, which hints a possible common
origin. 
Below $55 mK$ the fit is not as good. In this temperature 
range the $XY$ order parameter is big (larger than $1/2$) 
so there is no reason to expect $\zeta \sim m$.
However, $\zeta$ strikingly
saturates (unlike $m$ which shows the characteristic linear temperature
behavior of classical spin wave theory) so the discrepancy may be
indicating that quantum mechanical effects (or disorder) 
are important at low $T$.
\begin{figure}
\begin{center}
\leavevmode
\noindent
\epsfxsize=3.0in
\epsfysize=2.5in
\epsfbox{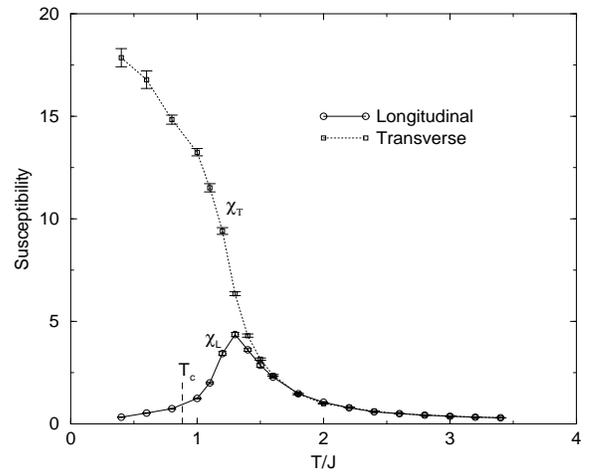}
\end{center}
\caption
{Longitudinal and transverse susceptibilties of a classical
nematic on a $100 \times 100$ lattice, for $h=0.05 J$.}
\label{fig:susceptibilities}
\end{figure}
Our analysis of the experiments strongly indicates that the 2DEG
in large magnetic fields in clean samples has regimes where it behaves like a
nematic fluid, an anisotropic metal. Such a metallic state should have a
strong signature in polarized light scattering experiments. In particular, a
nematic has long range fluctuations in the orientational order, which will
cause the polarization tensor correlation function (and the corresponding
longitudinal and transverse susceptibilities $\chi_L$ and $\chi_T$) to have a 
singularity
at $T_c$ (cut off by the anisotropy). This effect is similar to critical
opalescence but for orientational order instead of density fluctuations. For
non-zero and small background anisotropy $h$, for $T<T_c$, $\chi_L$ is
$\chi_L \sim h^{-\alpha}$, with $\alpha=1-1/\delta$, where
$\alpha=14/15$ at $T_c$. For $T>T_c$, $\chi_L$ can be written in a 
scaling form as $\chi(h,T)\sim \xi^{7/4}\Phi_{0}(h\xi^{15/8})$ where
$\Phi_{0}(0)=1$ and $\Phi_{0}(x) \sim x^{-14/15}$ as $x\rightarrow 
\infty$;  as discussed above, $\xi(T) \sim \exp(-A/\sqrt{t})$,
 where $t=T/T_c-1$.
Thus, at fixed but
small $h$, as the temperature is lowered, $\chi_L$ increases very rapidly 
to a maximum above $T_c$, with a crossover to 
a critical behavior $\sim h^{-\alpha(T)}$, where $\alpha(T)=2
(7+t)/(15+t)$ for $|h| \ll T$, and $\chi_L \sim (\pi T/2h)(1/(4\pi^2 J+h))$, 
for $T \ll |h|$. 
On the other hand, by Goldstone's Theorem, $\chi_T = m/h$, where $m$ is
the order parameter (Fig.\ \ref{fig:fit}). $\chi_L$ and $\chi_T$ are shown in
Fig.\ \ref{fig:susceptibilities}. Although the
specific heat of the 2DEG is very hard to measure, it should have a 
broad bump at a temperature $T^*>T_c$ set by the core energy of the
disclinations of the nematic phase, while at $T_c$ there is a very weak
essential singularity~\cite{KT} (rounded by the anisotropy).

Finally, we used the following procedure to extract the resistivities 
from the measured resistances. It was observed
recently~\cite{simon,19b} that much of the discrepancy 
between the data from Hall bars and square samples can be understood 
to be a consequence of
a large distortion of the current
distribution in square samples. If one assumes that the 2DEG has an 
anisotropic but homogeneous resistivity tensor~\cite{comment2}, one can 
calculate the distribution of currents using the method of conformal 
mappings~\cite{conformal}.  One finds that, for
an anisotropic homogeneous sample, with its principal axes aligned with the edges, 
\begin{eqnarray}
R_{xx}/R_{yy}=g(x)/g(1/x)
\label{eq:resistance_ratio}
\end{eqnarray}
where $x=  [L_{y} / L_{x}] \sqrt{{\rho_{yy}} / {\rho_{xx}}}\equiv x(T)$
measures the aspect ratio $L_y/L_x$
and the ratio of {\sl resistivities}, and
$g(x)$ is given by
\begin{eqnarray}
g(x)=\ln({{\theta_3(i\pi x/2) + \sqrt{k} \theta_2(i \pi x /2) } 
\over  {\theta_3(i\pi x /2) - \sqrt{k} \theta_2(i \pi  x /2) }})
\label{eq:g}
\end{eqnarray}
$\theta_2(z)$ and $\theta_3(z)$ are  
theta-functions with modulus $k$
\begin{equation}
k = 4  \sqrt{q} \prod_{n=1}^{\infty} ({{1+q^{2n}} \over { 1 + q^{2n-1}}})^4
\end{equation}
where $q= \exp(-2 \pi x)$ is the period ~\cite{conformal}. 
Given
$R_{xx}/R_{yy}$ at different temperatures,
and using Eq. (\ref{eq:resistance_ratio}),  we calculated the
function $x(T)$ (shown in the inset of 
Fig.\ \ref{fig:resistivities}).
At high $T$, $x(T)$ approaches a value somewhat
larger than $ 1$, but it is smaller than $1$ at lower temperatures, and both the 
resistances and the resistivities show a crossing at some high 
temperature~\cite{comment}. 
This effect indicates that the sample is not homogeneous at large
scales.
A macroscopic inhomogeneity is equivalent to an effective aspect 
ratio $L_y/L_x\neq 1$, and by choosing a value for $L_y/L_x=1.12$ we can make the 
ratio $\rho_{xx}/\rho_{yy}$ approach unity at high temperature. 

We thank J.\ P.\ Eisenstein and R.\ R.\ Du for making their data available us, 
and we thank J.\ P.\ Eisenstein, V.~J.~Emery, P.\ Goldbart, D.~Kivelson,
C.~Knobler, L.\ Radzihovsky, and M.\ Stone 
for useful discussions.
This work was supported in part by the NSF, grant numbers DMR98-08685 
at UCLA (SAK), NSF DMR98-17941 at UIUC, by the J.\ S.\ Guggenheim Foundation 
(EF), and by NASA grant No.\ NAG3-1841 at FSU (EM and KN).


\begin{references}

\bibitem{paper1} 
E.~Fradkin and S.~A.~Kivelson, {\prb} {\bf 59}, 8065 (1999).
\label{ref:paper1}

\bibitem{topo}  
S.~A.~Kivelson and V.~J.~Emery, 
{\it Synthetic Metals}, {\bf  80}, 151-158 (1996), and references therein.
\label{ref:topo}

\bibitem{nature} S.~A.~Kivelson,  E.~Fradkin and  V.~J.~Emery,
Nature {\bf 393}, 550 (1998).
\label{bib:nature}


\bibitem{FPA} H.\ Fukuyama, P.\ M.\ Platzman and P.\ W.\ Anderson, {\prb}
{\bf 19}, 5211 (1979).
\label{ref:FPA}

\bibitem{KFS} A.\ A.\ Koulakov, M.\ M.\ Fogler and B.\ I.\
Shklovskii, {\prl} {\bf 76}, 499 (1996).
\label{ref:FKS}

\bibitem{MC} R.\ Moessner  and J.\ T.\ Chalker, {\prb} {\bf 54}, 5006 (1996).
\label{ref:MC}

\bibitem{SMP} T.\ Stanescu, I.\ Martin and P.\ Phillips, cond-mat/9905116
\label{ref:SMP}

\bibitem{rezayi} E.\ H.\ Rezayi, F.\ D.\ M.\ Haldane and K.\ Yang, 
cond-mat/9903258.
\label{ref:rezayi}

\bibitem{vadim} V.\ Oganesyan {\sl et.\ al.\/}, in preparation.
\label{ref:vadim}

\bibitem{macdonald} T.\ Jungwirth, A.\ H.\ MacDonald, L.\ Smer{\u c}ka and 
S.\ M.\ Girvin, cond-mat/9905353.
\label{ref:macdonald}

\bibitem{caltech} M.\ P.\ Lilly, K.\ B.\ Cooper, J.\ P.\ Eisenstein, L.\
N.\ Pfeiffer and K.\ W. West, {\prl} {\bf 82}, 394 (1999).
\label{ref:caltech}

\bibitem{princeton} R.\ R.\ Du, H.\ St{\"o}rmer, D.\ C.\ Tsui, L.\
N.\ Pfeiffer and K.\ W. West, Solid State Comm.\ {\bf 109}, 389 (1999).
\label{ref:princeton}

\bibitem{2Dsmectics} D.\ R.\  Nelson and J.\ Toner, {\prb} {\bf 24}, 363 (1981). 
\label{ref:2Dsmectics}

\bibitem{tilt} W.\ Pan {\sl et.\ al.} , cond-mat/9903160; M.\ P.\ Lilly 
{\sl et.\ al.} , cond-mat/9903196.
\label{ref:tilt}

\bibitem{19b} M.\ P.\ Lilly, {\sl et.\ al.\/}, cond-mat/9903153.
\label{ref:19b}

\bibitem{caltech-unpublished} M.\ P.\ Lilly, {\sl et.\ al.\/} , unpublished.
\label{ref:caltech-unpublished}


\bibitem{KT} J.\ M.\ Kosterlitz and D.\ J.\ Thouless, {\jpc} {\bf 6},
1181 (1973).
\label{ref:KT}

\bibitem{fertig} H.\ Fertig, {\prl}{\bf 82}, 3693 (1999).
\label{ref:fertig}

\bibitem{schultka} N. Schultka and E. Manousakis, {\prb} {\bf 49}, 12071 (1994).
\label{ref:schultka}

\bibitem{comment3}The actual
calculation of this function requires a theory of transport in the nematic state.

\bibitem{simon} S.\ H.\ Simon, cond-mat/9903086; C.\ Wexler, unpublished.  
\label{ref:simon}

\bibitem{comment2} 
The longitudinal and Hall currents are mixed and the sample is not
homogeneous on large scales.
\label{ref:comment2}

\bibitem{conformal} Z.\
Nehari, {\sl Conformal Mapping}, Dover, New York (1952).
\label{ref:conformal}

\bibitem{caveat}  Even for small $m$, there is generally a
proportionality constant relating $m$ to $\zeta$, which we
set equal to unity so as to reduce the number of free parameters.

\bibitem{comment} This effect is absent in the data for higher Landau levels.





\end{references}
\end{document}